\documentclass[11pt]{article}

\usepackage[utf8]{inputenc} 
\usepackage[T1]{fontenc}    
\usepackage{hyperref}       
\usepackage{url}            
\usepackage{booktabs}       
\usepackage{amsfonts}       
\usepackage{nicefrac}       
\usepackage{microtype}      
\usepackage{lipsum}         
\usepackage{graphicx}       
\usepackage[numbers, sort&compress]{natbib}         
\usepackage{doi}            

\usepackage{amsmath}
\usepackage{amssymb}

\usepackage[margin=1in]{geometry}

\usepackage{listings}
\usepackage{fontawesome}

\usepackage{textcomp}
\usepackage{float}

\usepackage{color}

\usepackage{array}

\definecolor{dkgreen}{rgb}{0,0.6,0}
\definecolor{gray}{rgb}{0.5,0.5,0.5}
\definecolor{mauve}{rgb}{0.58,0,0.82}

\title{{\tt static\_maps}: {\tt consteval} {\tt std::map} and {\tt std::unordered\_map} Implementations in {\tt C++23}}

\author{
  Isaac D. Myhal\\
  Hillsdale College\\
  Dept. of Math and Computer Science
  \and
  Oliver Serang\\
  Hillsdale College\\
  Dept. of Math and Computer Science
}

\date{} 

\begin{document}
\maketitle

\begin{abstract}
\noindent Using {\tt consteval} from {\tt C++23}, we implement
efficient, new versions of {\tt std::map} and {\tt
  std::unordered\_map} for use when the keys are known at compile
time. We demonstrate superior performance of our {\tt unordered\_map}
on three demonstration use-cases: Lookup of elemental mass from atomic
symbol, lookup of amino acid from codon, and modification of stock
prices from S\&P 500 ticker symbols all produced runtimes $<40\%,
<35\%, <73\%$ of the respective runtimes of the {\tt std}
implementations. Our library runimes were $<80\%, <45\%, <97\%$ of the
lookup time of {\tt Frozen}, an alternative perfect hashing
implementation in {\tt C++} for problems also using {\tt constexpr}
keys. To our knowledge, this makes our library the overall fastest
drop-in (\emph{i.e.}, with a similar API) alternative to {\tt
  std::unordered\_map}. On one arbitrarily chosen demo, we demonstrate
runtimes $<35\%$ of {\tt PTHash} and $<89\%$ {\tt gperf},
state-of-the-art but not drop-in hashing libraries via external tools.
\end{abstract}

\noindent \textbf{Keywords:} data structures, {\tt C++}, hashing, perfect hashing

\section{Introduction}
\subsection{Mapping}
Mapping members of a collection keys $\mathcal{K}$ to their
corresponding values $\mathcal{V}$ is a fundamental task in computer
science. Where a naive approach performs lookup by comparing against
all keys in $O(n)$, ``ordered mapping'' uses a balanced binary tree to
guarantee $O(\log(n))$ lookup and insertion on a collection of
$n=|\mathcal{K}|$ keys \cite{adelson1962algorithm, bayer1972symmetric,
  guibas1978dichromatic, cormen2022introduction}. Red-black trees are
currently the prevalent implementation used for {\tt std::map}
\citep{cppref:map}. In practice, users of ordered maps on a fixed
collection of keys sometimes implement by sorting an array of keys and
searching with binary search rather than the greater overhead
(\emph{e.g.}, pointers to two child nodes) of the dynamic trees
specified by AVL and red-black trees.

\subsection{Hashing}
Hashing uses a function mapping keys to a codomain in $\mathbb{Z}$. In
doing so, hashing algorithms produce several bits at a time (using
parallelism in the processor's ALU) rather than the single bit
achieved by a compare in a balanced binary search tree. We will denote
this ``raw'' hash of the key as $h_{raw}(x), x\in \mathcal{K}$. These hashes can
use an integral {\tt SIZE\_TYPE} of 64, 32, 16, or 8 bits, but in
practice, users generally default to {\tt std::size\_t}, which is
specified as $\geq 16$ bits \cite{cppref:size_t};
however, in practice, x86-64 implementations commonly use
64-bits. These raw hashes are then mapped to $\mathbb{Z}/r \mathbb{Z}$
via $h_{table}$, where $r$ is the table size. \emph{E.g.}, a simple
implementation would use $h_{table}(i) = i \mod{r}$. Whether by
collision of the raw hashes (which is not expected if $h_{raw}$ is
designed well and {\tt SIZE\_TYPE} sufficiently large) or by collision
when mapping these larger integers to a contiguous table, collisions
at lookup are resolved by comparing keys themselves: $x,y\in \mathcal{K}$ will
be compared to one another if $h_{table}(h_{raw}(x)) =
h_{table}(h_{raw}(y))$ in order to retrieve the appropriate value.

Hashing is the basis of {\tt std::unordered\_map} (formerly {\tt
  hash\_map}), so called because it no longer stores the keys in
sorted order as a binary search tree
would \cite{cppref:unordered_map}. Performance of hash tables is
dependent on the number of collisions produced and on the efficiency
of the implementation.

\subsection{Perfect hashing}
Perfect hashing algorithms guaranteed implementations of $h_{table}$
that would result in no collisions (and thus no need to compare keys
themselves) as long as raw hashes were unique. The first linear-time
(to be clear, it was expected linear-time) algorithm with constant
lookup time uses two stages to construct $h_{table}$, each with hash
functions of the form $g_k(i, r) = k\cdot i \mod p \mod r$ with seed
parameter $k$ \cite{Fredmanetal1984}. Raw hash $i$ is mapped to a cell
in the primary table via
\[ g_{k^{(primary)}}(i,r^{(primary)}).\]
Values that collide in the primary table are hashed into a secondary
table composed only of those values with a primary table hash
collision (so several such secondary tables are made).

Here we slightly generalize \citeauthor{Fredmanetal1984}'s result for
practical efficiency \cite{serang:algorithms}. If the table size $r =
\delta\cdot n$, the probability that a randomly chosen $k$ would
produce a small total number of collisions $C$ can be bounded above by
$\frac{n\cdot\tau}{2\delta}$ using Markov's inequality:
\begin{eqnarray*}
\text{Pr}(C \geq \tau\cdot\mathbb{E}C) &\leq& \frac{1}{\tau}\\
1 - \text{Pr}(C \ge \tau\cdot\mathbb{E}C) &\geq& 1 - \frac{1}{\tau}\\
\text{Pr}(C < \tau\cdot\mathbb{E}C) &\geq& 1 - \frac{1}{\tau}\\
\end{eqnarray*}
And $\mathbb{E}C \leq \frac{n}{2\delta}$ (generalizing
\citeauthor{Fredmanetal1984}), thus $\tau\cdot\mathbb{E}C \leq
\frac{\tau\cdot n}{2\delta}$. So,
\[ \Pr\left(C < \frac{\tau\cdot n}{2\delta}\right) \geq 1 - \frac{1}{\tau}, \]
which is sufficient to find $k$ by random search until the collisions
are low. The expected number of such independent trials before success
(via the geometric distribution) is
\[\frac{1}{1 - \frac{1}{\tau}}.\]
Thus, we can tune $\delta$ (larger $\delta$ leads to fewer primary
table collisions and a lower chance of visiting the secondary table)
and $\tau$ (smaller $\tau$ leads to fewer collisions, but increases
the expected number of $k$ tried before finding a $k$ achieving the
bound) for practical performance while maintaining competitive
compilation times (where $k$ is chosen).

After finding a successful $k$ in the expected runtime from the
geometric, We can write $C$ as the sum of cells hashed into each
primary bucket choose 2:
\[ \sum_{j < r^{(primary)}}\binom{n_j}{2} = C < \frac{\tau\cdot n}{2\delta}. \]
This implies that for any constants $\tau,\delta$, at most
$O(\sqrt{n})$ raw hashes may be hashed into the same primary table
index; otherwise, we'd get a contradiction, because $\omega(n) >
\frac{\tau\cdot n}{2\delta}$ collisions would ensue. We can allocate
quadratic space to each secondary table $j$ to guarantee a high
probability of finding a $k^{(secondary)}_j$ producing no secondary
table collisions while keeping the total secondary table sizes $\in
O(n)$.

A more recent scheme, hash-displace-compress, also creates a primary
table with a primary hash. The primary table refers lookups to a
secondary hash function. The secondary hash function yields the
location of the value.  Because buckets can share the same secondary
hash, the memory usage may be less than the
\citeauthor{Fredmanetal1984} algorithm. Also, the values are
distributed through the whole of the second table, which uses memory
efficiently but could fail to find values that produce no collisions
in reasonable time \cite{belazzougui2009hash}.

\subsection{{\tt consteval} maps}
Here we implement a {\tt consteval} ordered {\tt map} by automating
the conversion of a fixed collection of keys from a binary search tree
to a sorted array. We also implement {\tt unordered\_map} with
compile-time perfect hashing.

Both implementations use the new {\tt consteval} qualifier from {\tt
  C++23}, which specifies a routine must run at compile time
\cite{cppref:consteval}. This allows {\tt constexpr} lookup of keys,
meaning keys known at compile time will have lookup optimized out
while still supporting lookup of keys unknown at compile time. The
library delivers state-of-the-art performance on three demonstration
problems: looking up atomic mass from atomic symbol, looking up amino
acid from codon, and modifying share price using a stock's ticker
symbol.

In addition to {\tt std::unordered\_map}, we compare to three hashing
libraries: {\tt Frozen} \cite{frozen2021grudinin} (an alternative
drop-in replacement for {\tt std::unordered\_map}), {\tt
  PTHash} \cite{pibiri2021pthash} (a {\tt C++} library), and {\tt
  gperf} \cite{schmidt1990gperf} (an external tool that accepts keyset
$\mathcal{K}$ and generates {\tt C++} code defining a hash
function). We demonstrate superior performance to each of these
libraries.

\section{Methods}
\subsection{Implementation}
Our {\tt map} implementation is straightforward, using the {\tt
  constexpr std::sort} implementation to sort a collection of
key-value pairs at compile time. Lookup is performed via {\tt
  std::lower\_bound} on the sorted array of keys. 

From here on, we focus on our implementation of {\tt unordered\_map}.

\subsubsection{Secondary tables and {\tt RaggedArray}}
The perfect hashing algorithm described above produces a table of
tables. The primary table is indexed by primary hash values and uses
$k^{(primary)}$. Each secondary table $j$ (where $j$ is the index in
the primary table to which this secondary table corresponds) has its
own secondary hash parameter $k^{(secondary)}_j$. The secondary tables
may have heterogeneous sizes. This could be implemented via {\tt
  std::vector<std::vector<T> >}; however, as of {\tt C++23}, {\tt
  std::vector} can be used as a temporary in {\tt consteval} contexts,
but not as a class member because it uses dynamic memory
allocations. Alternatively, this could be implemented via {\tt
  std::array<std::array<T,N2>, N1>}; however, keeping type homogeneity
of array elements would require {\tt N2=n} and {\tt N1 = r\_primary >
  n}, thus requiring $\Omega(n^2)$ space. Instead, we implement this
via a {\tt RaggedArray} class, which uses an array of the cumulative
sums of secondary table sizes used to compute an offset in another
array of size $r^{(primary)}$. 

Once an acceptable $k^{(primary)}$ is found and the {\tt RaggedArray}
is constructed, we find all $k^{(secondary)}_j$ values and store each
in an array {\tt std::array<SIZE\_T,N\_ROWS> ks\_secondary\_}. The
sizes of the secondary tables used to construct the {\tt RaggedArray}
are determined once we find a $k^{(primary)}$ achieving a total number
of collisions $C$ achieving the bound. Secondary table $j$ has a size
quadratic in $n_j$. As part of computing the total number of
collisions, we already compute $n_j$, the number of elements with
primary hash $j$. These $n_j$ values can be used to construct the {\tt
  constexpr RaggedArray}.

We group the keys that map to the same primary hash value before
finding the secondary hashes. We accomplish this with a temporary
vector, passed exclusively between {\tt consteval} functions. In
earlier versions, to eschew this temporary {\tt vector} in a {\tt
  consteval} context, we reordered the keys (and the associated
values) so that the same primary hash value were contiguous in $O(n)$
by using a helper array storing the first index for that primary hash
value (as with the values array) and incrementing these indices when
elements were placed in the corresponding array slice.

\subsubsection{Class hierarchy}
Perfect hashing eliminates the need to store keys for collision
resolution.  This can reduce the memory footprint of a hash table,
especially in cases where the values take little space. Nonetheless, a
program may iterate over the keys or values in which case the hash
table can encapsulate this data.  {\tt std::unordered\_map} provides
functions for retrieving the keys or values, so our maximal {\tt
  unordered\_map} provides those features using the most generally
applicable method. For increased performance in common cases, we
provide alternative classes with restricted functionality. All three
classes use {\tt at()} to check the validity of key lookup and do not
check key validity when using {\tt []}. No classes result in undefined
behavior or out of bounds memory access if {\tt at()} is used for lookup;
however, lookup of keys $x\not\in \mathcal{K}$ may result in retrieval of
a valid value instead of an error for some of the classes. When using
{\tt []}, none of the classes check if the key $x\in \mathcal{K}$.

We use encapsulation rather than inheritance for cleaner organization
and to avoid {\tt vtable} lookup from the ABC design pattern.

{\bf {\tt DangerousUnorderedMap} (keys not stored, values stored
  directly within secondary table, non-iterable):} {\tt
  DangerousUnorderedMap} is our primary building block that implements
perfect hashing of keys to values. It is so named because element
access with {\tt []} can produce undefined behavior when lookups are
called with a key $x\not\in \mathcal{K}$. {\tt at()} verifies that the
primary hash bin isn't empty to prevent any out-of-bounds access of
memory. But if an invalid key $x\not\in \mathcal{K}$ is queried with
{\tt at()}, it may still collide with some valid key $y\in
\mathcal{K}$ in the map and retrieve a valid value for that invalid
key without raising a warning or error. This class does not allow
iteration over the keys or values. We use this class internally in the
other classes.

{\bf {\tt unordered\_map\_small\_values} (keys stored, values stored
  directly within secondary table, iterable):} For cases where the
values are small types, {\tt unordered\_map\_small\_values}
encapsulates {\tt DangerousUnorderedMap<K,V>}. This means the values
are stored directly within the secondary table rather than storing
integer indices (thus avoiding an array lookup). The keys and values
are also stored in the {\tt unordered\_map\_small\_values} class in
the order they were originally provided. This is used to make it
iterable, but results in storing all values twice (once in the
encapsulated {\tt DangerousUnorderedMap} member and once in the value
array). Lookup of an invalid key $x\not\in \mathcal{K}$ may return a
valid value silently. In cases where values are small, this avoids a
secondary lookup, but with fewer safeguards when looking up invalid
keys.

{\bf {\tt unordered\_map} (keys stored, value indices stored in
  secondary table, iterable):} For the general case, our {\tt
  unordered\_map} stores keys and values in their original order and
encapsulates a {\tt DangerousUnorderedMap<K,SIZE\_T>} member
internally to facilitate lookup; the output integral type is used to
index an array of values. The array of values also allows iteration
over keys and values and does not require values to be stored
twice. It also allows {\tt at()} to throw an exception if the key is
not in the map (by retrieving an invalid integer {\tt
  SIZE\_T(-1)}). Access with {\tt []} behaves like {\tt
  DangerousUnorderedMap::at()} with only a lightweight check that
there will be no out-of-bounds memory access (but the possibility of
returning a valid value when looking up an invalid key). The
indirection from the extra array lookup, {\tt unordered\_map} allows
the values to be mutable unless the {\tt unordered\_map} is declared
{\tt const} or {\tt constexpr}. If the {\tt unordered\_map} is
declared {\tt constinit}, it ensures compile-time construction but
allows the values to be changed.

{\bf {\tt unordered\_map\_mutable\_values} (keys stored, value indices
  stored in secondary table, iterable):} If {\tt constinit} cannot be
used for some reason and the use-case requires a non-const {\tt
  unordered\_map} (to allow modification of values after construction)
is not properly optimized by a particular compiler, {\tt
  unordered\_map\_mutable\_values} encapsulates a {\tt constexpr
  unordered\_map<K,SIZE\_T>} to map keys to indices and then use the
retrieved index to look up a mutable value in the encapsulated {\tt
  std::array<V,N> values}.

\subsubsection{{\tt consteval} generation of pseudorandom numbers}
Random number generators from the standard library cannot be called at
compile time, so we implemented a simple generator class exclusively
for compile time.  The algorithm is a 64-bit version of
xorshift \cite{vigna2016xorshift, handwiki_xorshift}. If no seed is
provided, we automatically generate a seed from the compile timestamp
using the standard {\tt \_\_DATE\_\_} and {\tt \_\_TIME\_\_} macros.

These random numbers are used to generate values for $k^{(primary)},
k^{(secondary)}$. 

\subsubsection{Fast compile-time modulo}
\lstset{language=C++,
   basicstyle=\ttfamily\footnotesize,
   keywordstyle=\color{blue}\ttfamily,
   stringstyle=\color{red}\ttfamily,
   commentstyle=\color{magenta}\ttfamily,
   morecomment=[l][\color{magenta}]{\#},
   breaklines=true,
   label=fast-modulo
}
\begin{lstlisting}[caption={Modulo operations optimized for the case where the divisor is near the maximum for the unsigned integer's type.}]
template <typename SIZE_T, typename PRIME_T>
constexpr SIZE_T mod_largest_prime(const SIZE_T n) {
  static_assert_all_possible_n_less_than_2_times_largest_prime<PRIME_T>();
  return n < largest_prime_of_type<PRIME_T>() ? n : n - largest_prime_of_type<PRIME_T>();
}

template <typename SIZE_T>
constexpr SIZE_T mod_largest_prime(const SIZE_T n) {
  return mod_largest_prime<SIZE_T, SIZE_T>(n);
}

template <typename SIZE_T, typename PRIME_T>
constexpr SIZE_T add_value_and_value_much_less_than_largest_prime_mod_largest_prime(SIZE_T a, SIZE_T b) {
  static_assert_all_possible_n_less_than_2_times_largest_prime<PRIME_T>();
  return largest_prime_of_type<PRIME_T>() - b <= a ?
    a + b - largest_prime_of_type<PRIME_T>() :
    a + b;
}

template <typename SIZE_T>
constexpr SIZE_T add_value_and_value_much_less_than_largest_prime_mod_largest_prime(SIZE_T a, SIZE_T b) {
  return add_value_and_value_much_less_than_largest_prime_mod_largest_prime<SIZE_T, SIZE_T>(a, b);
}

template <uint16_t EXPONENT, typename SIZE_T>
constexpr SIZE_T times_power_of_two_mod_largest_prime(const SIZE_T h) {
  static_assert(std::numeric_limits<SIZE_T>::max() - largest_prime_of_type<SIZE_T>() + 1 < std::numeric_limits<uint8_t>::max());
  static constexpr SIZE_T TWO_TO_64_MOD_P = std::numeric_limits<SIZE_T>::max() - largest_prime_of_type<SIZE_T>() + 1;
  static constexpr uint8_t HALF_BITS = std::numeric_limits<SIZE_T>::digits / 2;
  static_assert(TWO_TO_64_MOD_P <= std::numeric_limits<SIZE_T>::max() / 2);

  const SIZE_T h_hi = upper_bits(h);
  const SIZE_T h_lo = lower_bits(h);

  // NOTE: could be generalized
  if constexpr (EXPONENT == HALF_BITS) {
    return add_value_and_value_much_less_than_largest_prime_mod_largest_prime(
      shift_preserve_first_type(h_lo, HALF_BITS),
      multiply_preserve_type(h_hi, TWO_TO_64_MOD_P)
    );
  }
  else if constexpr (EXPONENT == HALF_BITS * 2) {
    return add_value_and_value_much_less_than_largest_prime_mod_largest_prime<SIZE_T>(
      times_power_of_two_mod_largest_prime<HALF_BITS, SIZE_T>(
        multiply_preserve_type(h_hi, TWO_TO_64_MOD_P)
        ),
      multiply_preserve_type(h_lo, TWO_TO_64_MOD_P)
    );
  }
  else {
    static_assert(false, "Unsupported exponent");
    return h;
  }
}

template <typename SIZE_T>
constexpr SIZE_T multiply_mod_largest_prime(const SIZE_T h, const SIZE_T k) {
  static_assert(!std::numeric_limits<SIZE_T>::is_signed);
  if constexpr (std::is_same_v<SIZE_T, uint64_t>) {
  #ifdef USING_UINT128
    double_bits_t<SIZE_T> product = uint_with_double_bitsize(h) * k;

    SIZE_T high_bits = upper_bits(product);
    product = lower_bits(product);
    product += times_power_of_two_mod_largest_prime<64, SIZE_T>(high_bits); // up to (2^64-1)60

    high_bits = upper_bits(product); // up to 60
    product = lower_bits(product);
    product += times_power_of_two_mod_largest_prime<64, SIZE_T>(high_bits); // up to (2^64-1) + 60*59 < 2^64-59 + 61 * 59

    return mod_largest_prime<double_bits_t<SIZE_T>,SIZE_T>(product);
  #else
    auto multiply_lower_or_upper_bits = [](const half_bits_t<SIZE_T> h_part, const half_bits_t<SIZE_T> k_part) -> uint64_t {
      // No need to use multiply_preserve_type since these are never short int or char
      return mod_largest_prime(static_cast<SIZE_T>(h_part) * k_part);
    };

    const SIZE_T z_low = multiply_lower_or_upper_bits(lower_bits(h), lower_bits(k));
    const SIZE_T z_middle = add_value_and_value_much_less_than_largest_prime_mod_largest_prime(
      multiply_lower_or_upper_bits(upper_bits(h), lower_bits(k)),
      multiply_lower_or_upper_bits(lower_bits(h), upper_bits(k))
    );
    // Don't need the mod for z_high, but this takes care of casting
    const SIZE_T z_high = multiply_lower_or_upper_bits(upper_bits(h), upper_bits(k));

    return add_value_and_value_much_less_than_largest_prime_mod_largest_prime<SIZE_T>(
      add_value_and_value_much_less_than_largest_prime_mod_largest_prime<SIZE_T>(
        times_power_of_two_mod_largest_prime<std::numeric_limits<SIZE_T>::digits, SIZE_T>(z_high),
        times_power_of_two_mod_largest_prime<std::numeric_limits<SIZE_T>::digits / 2, SIZE_T>(z_middle)
      ),
      z_low
    );
  #endif
  }
  else {
    const double_bits_t<SIZE_T> product = uint_with_double_bitsize(h) * k;
    // When doing the naive method, the product could be larger than
    // 2 * largest_prime, so we cannot use the specialized mod_largest_prime
    return static_cast<SIZE_T>(product % largest_prime_of_type<SIZE_T>());
  }
}
\end{lstlisting}

We implement a custom algorithm for multiplying two numbers modulo a
large prime. This algorithm leverages the standard behavior of
unsigned 64-bit integers wrapping around to 0 at $2^{64}$. Because our
chosen primes are close to the maximum integer that can be stored in
64 bits, a modulo can be accomplished by a comparion and possible
subtraction faster than the standard modulo operator {\tt \%}. When
multiplying $a\cdot b \mod{P}$, the wraparound behavior is equivalent
to a modulo operation by $2^{64}$, so we can avoid using the modulo
operator again. The adjustment is a multiple of $2^{64} - P$,
where $P$ is our large prime. A similar operation can be done with
smaller sizes of integers as well.

\subsubsection{Flexible {\tt SIZE\_TYPE}}
Because many applications of hash maps have $|\mathcal{K}| = n \ll
2^{32}$ keys, it can be worthwhile to use smaller integer types than
{\tt std::size\_t}. Using smaller integer types can also increase the
throughput of SIMD parallelism for the hashing process. In particular,
using 32-bit integers instead of 64-bit integers allows for the hash
function to perform a multiplication without handling numbers larger
than the 64-bit integer limit. Furthermore, a bucket size must be
stored for every value of the primary hash, so using a smaller integer
type allows more of that array to fit in cache at one time. The
constructor determines the size type by default from the number of
keys.

If 64-bit indices are needed, multiplication is handled by our custom
modulo functions to avoid integer overflow. By applying our modulo
operation to intermediate steps of the multiplication, we avoid the
need to store a 128-bit integer and the use of the {\tt \%}
operator. We also provide the option to use a 128-bit integer type to
perform this multiplication in case the library user has a fast
implementation in hardware. This can be enabled at compilation with
{\tt -D~USING\_UINT128}. The compiler-specific name of the 128-bit
integer type can be set in {\tt uint128.hpp}.

\subsubsection{Avoidance of secondary hashing}
When only one key is found for a primary hash, {\tt unordered\_map}
can simply return element 0 of the appropriate secondary table (which
will have size 1), thereby avoiding the need of secondary
hashing. This speedup is more pronounced when few collisions occur in
the primary table. This can be achieved by increasing $\delta$,
although using too large a $\delta$ will stress the CPU cache more.

\subsubsection{Factory construction}
We allow customization of the size type and the space allocated to the
tables. We also provide a feature called {\tt USE\_SALT} that changes
the values from the raw hash function. No two distinct values can
produce the same value when each XORed with a fixed value; therefore,
XOR with a fixed salt produces a random shuffling of raw hashes. For
the cost of a single XOR (1 clock cycle on many modern CPUs), a lower
secondary hash frequency may be required. The best values for these
parameters would come from benchmarking on a specific keyset
$\mathcal{K}$. For a more convenient process, we provide a factory
function {\tt make\_unordered\_map} that can infer good parameter
values from the keyset $\mathcal{K}$.  The ordinary constructor must
determine allocation sizes for class members (\emph{e.g.}, arrays)
from template parameters, so they cannot depend on inferred parameters
using the existing {\tt C++} standards. {\tt make\_unordered\_map}
resolves this by choosing such parameters outside of the
constructor. Here the geometric bound on the expected time for finding
$k^{(primary)}, k^{(secondary)}$ can be used to time out and declare a
given {\tt SIZE\_TYPE} unsuccessful.

Specifically, this factory function applies the raw hash to the keyset
$\mathcal{K}$ and truncates the values to different sizes. It selects
minimum size where collisions do not occur.  In the event that any
values differ by exactly the large prime for that size type, we enable
{\tt USE\_SALT} to avoid a collision.

To use these functions, the data must be placed in a {\tt constexpr}
variable first for passing as a template argument:\\ {\tt
  static~constexpr~auto~data\_array~=
  std::to\_array<std::pair<key\_t,~value\_t>~>( \{\{key0,val0\}, ...\}
  );\\ auto~key\_to\_value~=~static\_maps::make\_unordered\_map<data\_array>();}\\ As
an added advantage, other template parameters (such as $n$, the number
of key-value pairs) will be inferred, so this method is more robust to
changes in the data. The initializer list of key-value pairs can be
imported from an external data source via {\tt \#include}.

One parameter that is not inferred is $\delta$, implemented via {\tt
  DELTA\_PRIMARY}. It controls the constant factor in the $O(n)$ size
of the table. We calculate the exact constant factor by rounding {\tt
  DELTA\_PRIMARY} up to the nearest power of two. Rounding to a power
of 2 makes for more efficient modulo by bitmasking or casting to the
appropriate integral type. The $O(n)$ bound guarantees that some
factor will work in a reasonably short compilation but does not
suggest what the factor should be in practice. The effects of using
different values vary depending on the system, so we did not implement
an algorithm to infer its value.

\subsubsection{Drop-in API}
Where a {\tt std::unordered\_map} is already being used with a static
keyset $\mathcal{K}$, the only code change required is to change the
namespace from {\tt std} and add the number of keys to the template
parameters (after value type {\tt T}). If the {\tt KeyEqual} or {\tt
  Allocator} optional template parameters are being set, our code will
not work without modification.

Lookup speed may be increased by tuning the template parameters
manually. Enabling {\tt USE\_SALT} may lower collisions and improve
cache performance, but it adds the overhead of an XOR operation during
hashing (1 clock cycle on many modern CPUs)). {\tt DELTA\_PRIMARY}
controls how big the primary table is. Decreasing {\tt DELTA\_PRIMARY}
generally increases compilation time and decreases memory footprint,
which can improve cache performance. To modify {\tt
  R\_SECONDARY\_TOTAL}, derive its value from {\tt
  get\_secondary\_table\_total\_size\_bound(N,~DELTA\_PRIMARY,~tau)}
with your chosen $\tau$. That function is in {\tt
  calculate\_table\_size.hpp}. Lower $\tau$ values generally increase
compilation time but decrease the number of keys which need to run the
secondary hash function. Values of $\tau\leq 1$ could make compilation
infeasible, but a value just above $1$ tends to work well. Note that
$\tau$ could not be its own template parameter because it is a
floating point number.

\subsubsection{{\tt caching} the {\tt unordered\_map} before low-latency access}
Because construction is done at compile time, the hash table will be
found in the {\tt .rodata} section and be immutable. This has
advantages (neither allocations nor time for construction at runtime);
however, before low-latency use of the {\tt []} operator, recently
having touched the keys and values is advantageous. To this end, our
library provides a {\tt cache()} member function which iterates
through all keys and looks up the corresponding value at runtime. To
prevent it from being optimized out, a total is computed over the
values in a {\tt volatile double} value (an integral type risks being
optimized out). This can prepare a {\tt unordered\_map} immediately
before low-latency access is required, increasing hits to the CPU
cache.

\subsection{Runtime experiments}
An {\tt unordered\_map} can also be used to lookup keys based on
some input at runtime. Having fixed keys lets lookups skip collision
detection, since we choose a perfect hash during compilation. So,
lookups take less time while the program is running. These demos focus
on speed gains when performing access operations at runtime. 

Note that the same {\tt unordered\_map} could be used to look up a
collection of constants at compile time for readability. For code that
is already doing this, switching to {\tt static\_maps::unordered\_map}
would immediately decrease runtime. When the key is known at compile
time, the lookup is performed during compilation and the value is
placed as a literal in the resulting program. There is then zero
runtime impact of using a lookup.

\subsubsection{Chemistry demo: lookup of average isotopic mass from atomic symbol}
Our benchmark during development was a mapping from short strings
(element abbreviations) to floating point numbers (average atomic mass
for that element). Such a map might be found in chemistry or biology
software \cite{kreitzberg2020fast}. The keyset and values remain
constant, while each key used for lookup is a runtime variable.

\lstset{language=C++,
   basicstyle=\ttfamily\footnotesize,
   keywordstyle=\color{blue}\ttfamily,
   stringstyle=\color{red}\ttfamily,
   commentstyle=\color{magenta}\ttfamily,
   morecomment=[l][\color{magenta}]{\#},
   breaklines=true,
   label=element-to-mass-demo
}
\begin{lstlisting}[caption={Isotopic mass indexed by element}]
#include "hashes.hpp"
// make hash<K> default available for static_unordered_map:
using namespace constexpr_hashes;

#include "static_unordered_map.hpp"
#include <iostream>
#include <chrono>

int main() {
  // use string_view via ""sv.
  // this allows comparison of const char* by string:
  using namespace std::literals;

  std::cout << "seed " << seed() << "\n";

  auto t1 = std::chrono::high_resolution_clock::now();
  
  static constexpr auto data_array = std::to_array<std::pair<std::string_view, double>>({
    #include "data.txt"
  });
  static constexpr auto element_to_mass = static_maps::make_unordered_map<data_array>();
  element_to_mass.cache();

  auto t2 = std::chrono::high_resolution_clock::now();
  auto time_construction_ns = (t2 - t1).count();
  std::cout << "construction (ns) " << time_construction_ns << std::endl;

  srand(0);

  static constexpr std::size_t N=1<<10;
  // ignore first few runtimes in case sorting vectors, etc. for setup has an impact:
  static constexpr std::size_t N_IGNORE=0;

  double time_lookup_tot_ns = 0.;

  // to prevent dead code elim and to check random sequences are the same:
  double mass_total = 0.;

  // NOTE: static maps could use static array here, but use vector to
  // ignore impact of heap usage (even though eschewing heap usage
  // benefits static maps):
  std::vector<std::string_view> keys(element_to_mass.size());
  std::size_t j=0;
  for (const auto& p : element_to_mass) {
    keys[j] = std::get<0>(p);
    ++j;
  }
  // sort so that keys looked up are at same indices for all tests:
  std::sort(keys.begin(), keys.end());

  for (std::size_t i=0; i<N; ++i) {
    auto index = random() % element_to_mass.size();
    const auto& element = keys[index];

    auto t1 = std::chrono::high_resolution_clock::now();
    const auto& mass = element_to_mass[element];
    auto t2 = std::chrono::high_resolution_clock::now();

    if (i >= N_IGNORE)
      time_lookup_tot_ns += (t2 - t1).count();
    mass_total += mass;
  }
  std::cout << "lookup mass tot [to check consistency across methods] " << mass_total << std::endl;
  std::cout << "lookup time (ns) tot " << time_lookup_tot_ns << std::endl;
  std::cout << "lookup time (ns) avg " << time_lookup_tot_ns/(N-N_IGNORE) << std::endl;

  std::cout << "\nnet time: construction + lookup tot (ns) " << time_construction_ns + time_lookup_tot_ns << std::endl;
}
\end{lstlisting}

\subsubsection{Biology: lookup of amino acid from codon}
This demo maps strings of three letters to single amino acid
characters and uses this lookup to translate a string of codons into
amino acids.

\lstset{language=C++,
   basicstyle=\ttfamily\footnotesize,
   keywordstyle=\color{blue}\ttfamily,
   stringstyle=\color{red}\ttfamily,
   commentstyle=\color{magenta}\ttfamily,
   morecomment=[l][\color{magenta}]{\#},
   breaklines=true,
   label=element-to-mass-demo
}
\begin{lstlisting}[caption={Amino acid symbols indexed by codon}]
#include "hashes.hpp"
// make hash<K> default available for static_unordered_map:
using namespace constexpr_hashes;

#include "static_unordered_map.hpp"
#include <iostream>
#include <chrono>

int main() {
  // use string_view via ""sv.
  // this allows comparison of const char* by string:
  using namespace std::literals;

  // generate a dna string:
  char nucleotides[] = {'G', 'A', 'T', 'C'};
  unsigned int n = 1<<16;
  std::string gene(n*3, ' ');
  for (unsigned i=0; i<n*3; ++i)
    gene[i] = nucleotides[rand() % 4];

  std::string protein(n, '?');

  // build an unordered_map for codon to amino acid:
  auto t1 = std::chrono::high_resolution_clock::now();

  constexpr auto codon_to_amino = static_maps::unordered_map(std::to_array<std::pair<std::string_view, char>>({{"AAA",'K'},{"AAC",'N'},{"AAG",'K'},{"AAT",'N'},{"ACA",'T'},{"ACC",'T'},{"ACG",'T'},{"ACT",'T'},{"AGA",'R'},{"AGC",'S'},{"AGG",'R'},{"AGT",'S'},{"ATA",'I'},{"ATC",'I'},{"ATG",'M'},{"ATT",'I'},{"CAA",'Q'},{"CAC",'H'},{"CAG",'Q'},{"CAT",'H'},{"CCA",'P'},{"CCC",'P'},{"CCG",'P'},{"CCT",'P'},{"CGA",'R'},{"CGC",'R'},{"CGG",'R'},{"CGT",'R'},{"CTA",'L'},{"CTC",'L'},{"CTG",'L'},{"CTT",'L'},{"GAA",'E'},{"GAC",'D'},{"GAG",'E'},{"GAT",'D'},{"GCA",'A'},{"GCC",'A'},{"GCG",'A'},{"GCT",'A'},{"GGA",'G'},{"GGC",'G'},{"GGG",'G'},{"GGT",'G'},{"GTA",'V'},{"GTC",'V'},{"GTG",'V'},{"GTT",'V'},{"TAA",'$'},{"TAC",'Y'},{"TAG",'$'},{"TAT",'Y'},{"TCA",'S'},{"TCC",'S'},{"TCG",'S'},{"TCT",'S'},{"TGA",'$'},{"TGC",'C'},{"TGG",'W'},{"TGT",'C'},{"TTA",'L'},{"TTC",'F'},{"TTG",'L'},{"TTT",'F'}}));
  // NOTE: doesn't cache (should eventually get saturated anyway)

  auto t2 = std::chrono::high_resolution_clock::now();

  // translate codon to amino acid:
  for (unsigned i_aa=0; i_aa<n; ++i_aa) {
    auto codon = gene.substr(3*i_aa, 3);
    protein[i_aa] = codon_to_amino[codon];
  }

  auto t3 = std::chrono::high_resolution_clock::now();

  std::cout << protein.substr(0,10) << "..." << std::endl;

  auto time_ctor_ns = (t2 - t1).count();
  auto time_lookup_ns = (t3 - t2).count();
  auto time_ns = time_ctor_ns + time_lookup_ns;
  std::cout << "time (ns) " << time_ns << " = ctor " << time_ctor_ns << " + lookup " << time_lookup_ns << std::endl;
}
\end{lstlisting}

\subsubsection{Finance: modification of stock price from ticker symbol}
This demo times how long it takes to \emph{change} many values in the
{\tt unordered\_map}. This simulates a financial application where an
external source prompts changes in values for various keys. The
selection of which keys are changed each operation is random to
simulate data being streamed as stock price changes occur. In
practice, it may be more efficient to first map the symbols to unique
integer types (\emph{e.g.}, via an {\tt enum}) for processing with
simpler table lookup; however, even with such a trick, it would still
be necessary to translate {\tt const char*} ticker symbols to these
unique integer types.

\lstset{language=C++,
   basicstyle=\ttfamily\footnotesize,
   keywordstyle=\color{blue}\ttfamily,
   stringstyle=\color{red}\ttfamily,
   commentstyle=\color{magenta}\ttfamily,
   morecomment=[l][\color{magenta}]{\#},
   breaklines=true,
   label=element-to-mass-demo
}
\begin{lstlisting}[caption={Prices indexed by stock symbol, changing the values while the keyset stays constant}]
#include "hashes.hpp"
// make hash<K> default available for static_unordered_map:
using namespace constexpr_hashes;

#include "static_unordered_map.hpp"
#include "random.hpp"
#include "split.hpp"
#include <iostream>
#include <random>
#include <chrono>

int main() {
  // use string_view via ""sv.
  // this allows comparison of const char* by string:
  using namespace std::literals;

  // build an unordered_map for codon to amino acid:
  auto t_start = std::chrono::high_resolution_clock::now();

  static constexpr auto stock_data = std::to_array<std::pair<std::string_view, int>>({
    #include "stock_data.txt"
  });  
  static constinit auto stock_to_price = static_maps::unordered_map(stock_data);
  
  auto t_pause = std::chrono::high_resolution_clock::now();

  static constexpr std::array stock_symbols = std::get<0>(static_maps::split(std::to_array<std::pair<std::string_view, int>>({
    #include "stock_data.txt"
  })));

  std::mt19937 rand_eng(seed());
  std::normal_distribution rand_dist(0.0, 50.0);

  auto time_lookup = t_pause - t_pause;
  auto t_resume = std::chrono::high_resolution_clock::now();

  for (std::size_t i=0; i<64*stock_symbols.size(); ++i) {
    std::string_view symbol;
    std::ranges::sample(stock_symbols, &symbol, 1, rand_eng);
    double change = std::lround(rand_dist(rand_eng));

    auto t1 = std::chrono::high_resolution_clock::now();
    stock_to_price[symbol] += change;
    auto t2 = std::chrono::high_resolution_clock::now();
    
    time_lookup += t2 - t1;
  }

  auto t_end = std::chrono::high_resolution_clock::now();

  constexpr std::string_view sample = "MSFT"sv;
  auto cents = stock_to_price[sample] % 10000 / 100;
  std::cout << (stock_to_price[sample] / 10000) << '.' << (cents < 10 ? "0" : "") << cents << std::endl;

  auto time_ctor_ns = (t_pause - t_start).count();
  auto time_lookup_ns = time_lookup.count();
  auto time_ns = time_ctor_ns + time_lookup_ns;
  std::cout << "time (ns) " << time_ns << " = ctor " << time_ctor_ns << " + lookup " << time_lookup_ns << std::endl;
}

\end{lstlisting}

\subsubsection{Benchmark hardware, compiler, and compilation flags}
All demos were timed on a laptop. Specifically, they were run on an
Intel Core i5-1240P @4.4 GHz CPU (12 cores, 24 threads, 448 KiB/640
KiB/9 MiB/12 MiB of levels 1d/1i/2/3 cache). All demos were timed with
lowest nice value (\emph{i.e.}, {\tt nice -n -20}), near real-time
priority {\emph{i.e.}, {\tt chrt -f 98}), and ASLR disabled
  (\emph{i.e.}, {\tt setarch `uname -m` -R}). {\tt g++} version 15.2.1
  was used and compiled with {\tt -std=c++23 -O3 -march=native} flags.

For the element to atomic mass demo, the {\tt gcc} flag {\tt
  -fconstexpr-ops-limit=100000000} was needed to construct the {\tt
  static\_maps::unordered\_map} at compile time. For the stock symbol
to price demo, {\tt -fconstexpr-ops-limit=20000000000} was used. These
control the number of operations before {\tt consteval} functions are
deemed to have failed, similar to template recursion depth in earlier
versions of {\tt gcc}.

\section{Results}
\subsection{{\tt static\_maps::map} vs. {\tt std::map}}
\begin{figure}[H]
  \centering
\begin{tabular}{c>{\centering\arraybackslash}m{3in}}
  & vs. {\tt std::map}\\
  \rotatebox{90}{\textbf{Elements}} & \includegraphics[width=2.95in]{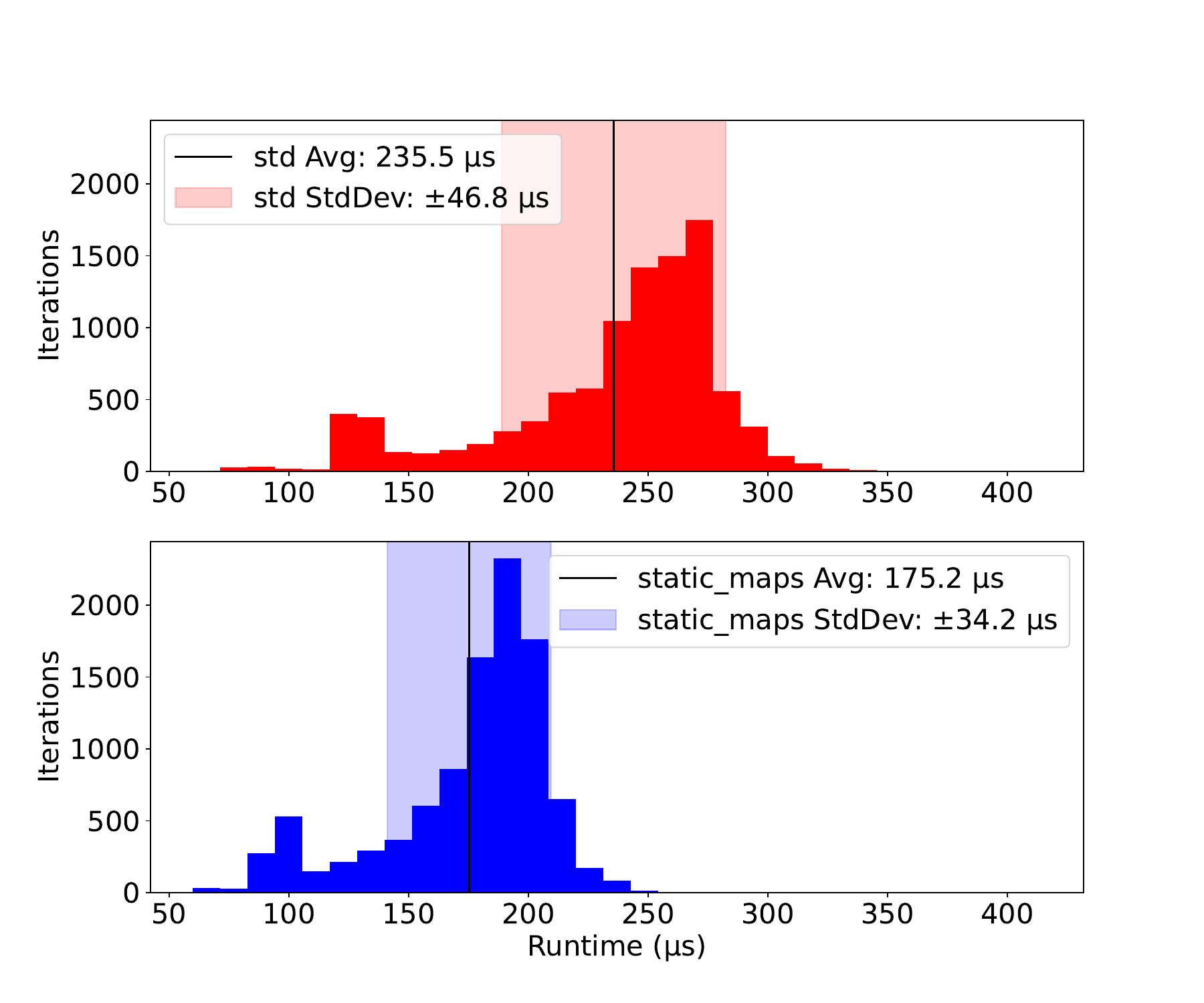} \\
  \rotatebox{90}{\textbf{Codons}} & \includegraphics[width=2.95in]{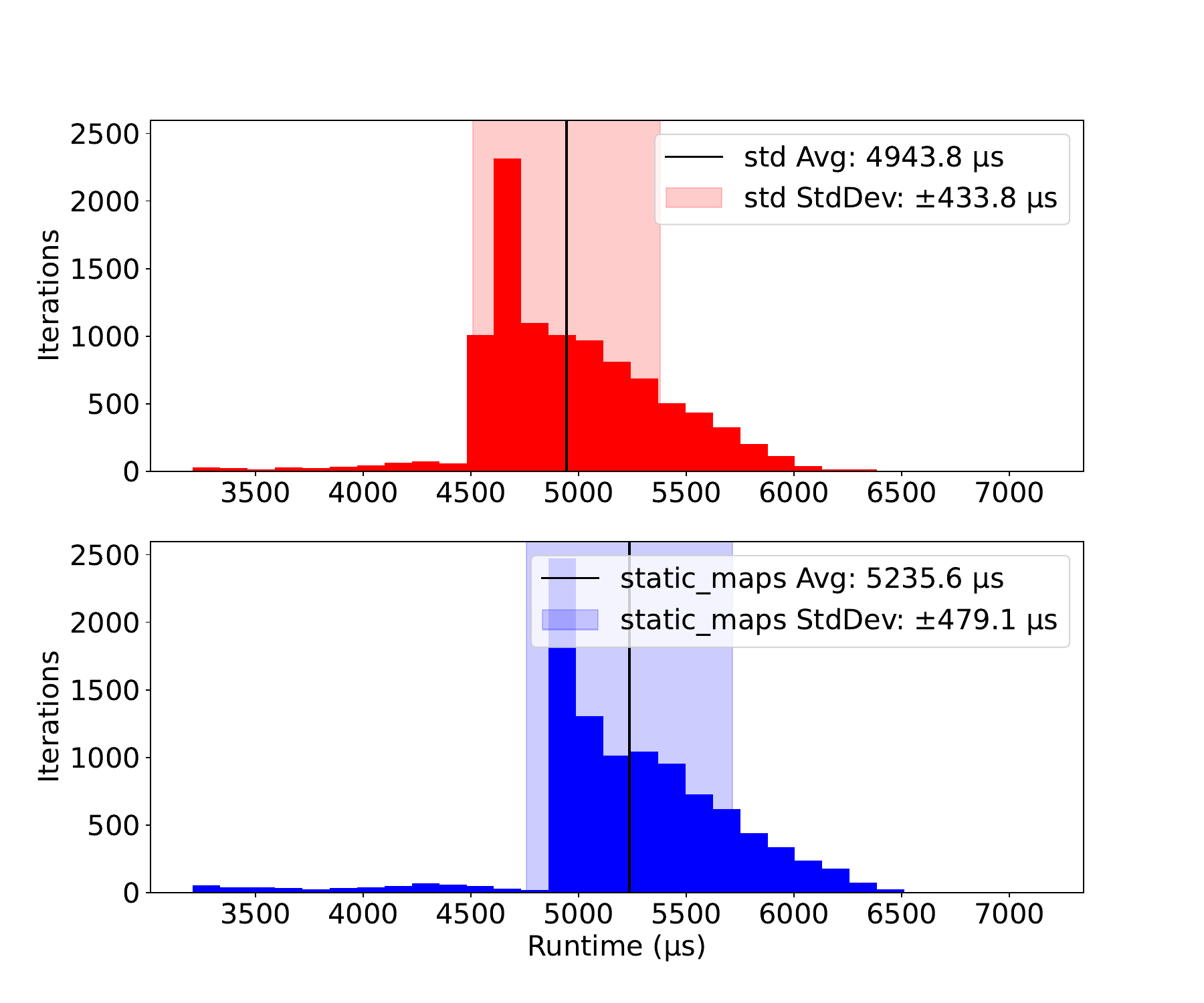} \\
  \rotatebox{90}{\textbf{Stocks}} & \includegraphics[width=2.95in]{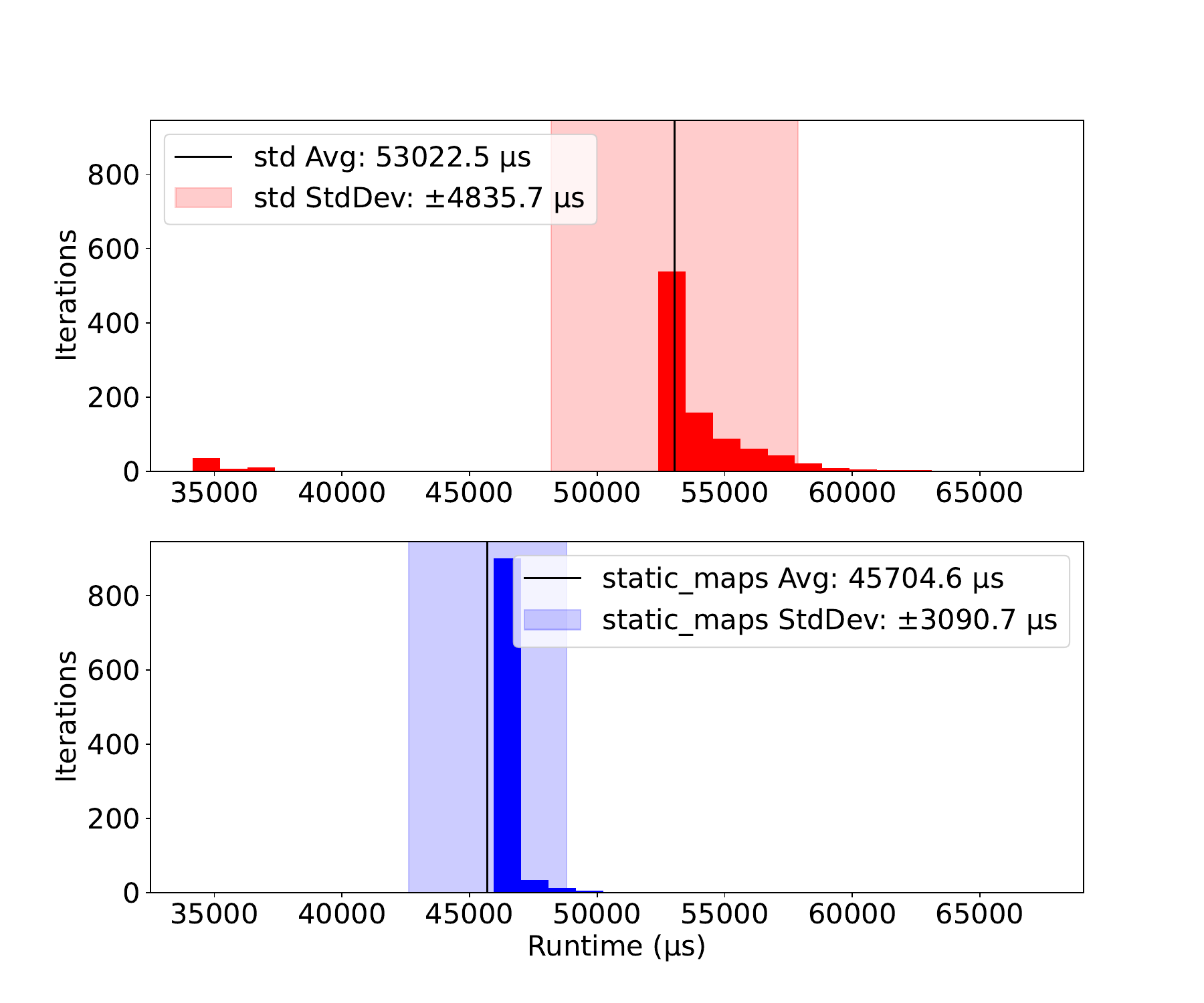}
\end{tabular}
\caption{Runtimes of construction and lookup for {\tt static\_maps::map} against {\tt std::map} on various demos. Lower is better.}
\label{fig:map-demos}
\end{figure}

\subsection{{\tt static\_maps::unordered\_map} vs. {\tt std::unordered\_map}, {\tt Frozen}}
\begin{figure}[H]
  \centering
  \begin{tabular}{c>{\centering\arraybackslash}m{3in}>{\centering\arraybackslash}m{3in}}
    & vs. {\tt std::unordered\_map} & vs. {\tt frozen::unordered\_map} \\
\rotatebox{90}{\textbf{Elements}} & \includegraphics[width=2.95in]{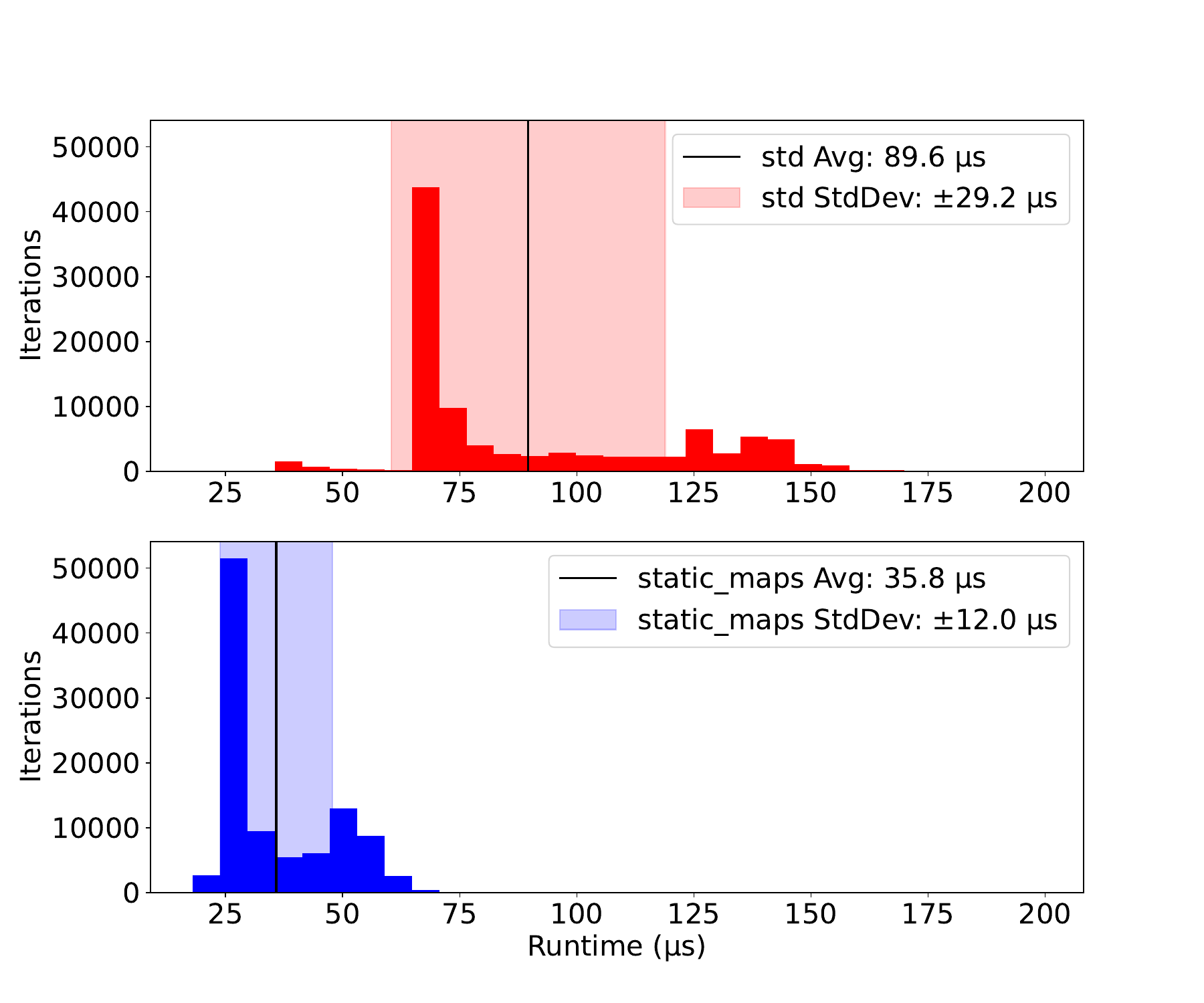} & \includegraphics[width=2.95in]{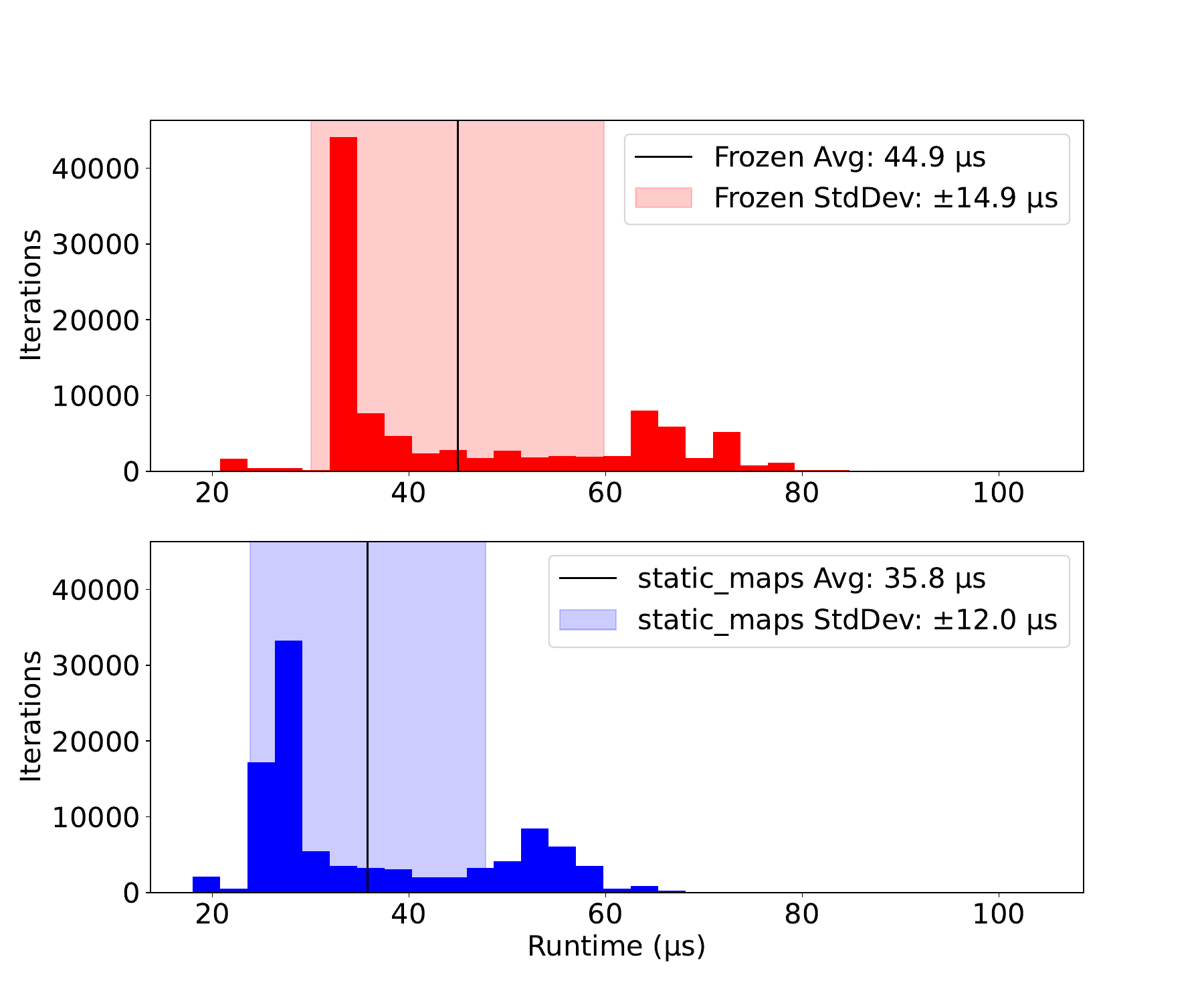} \\
\rotatebox{90}{\textbf{Codons}} & \includegraphics[width=2.95in]{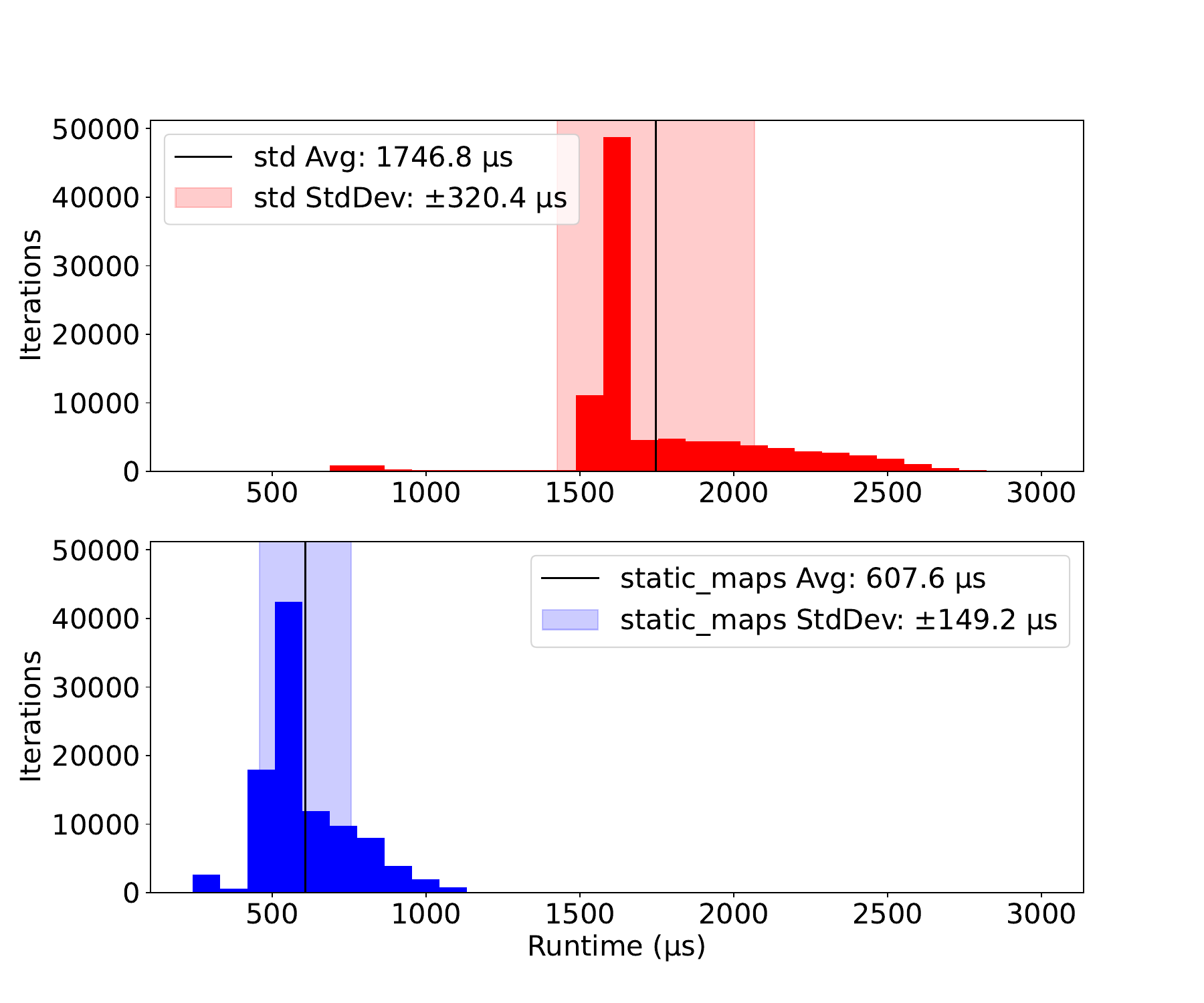} & \includegraphics[width=2.95in]{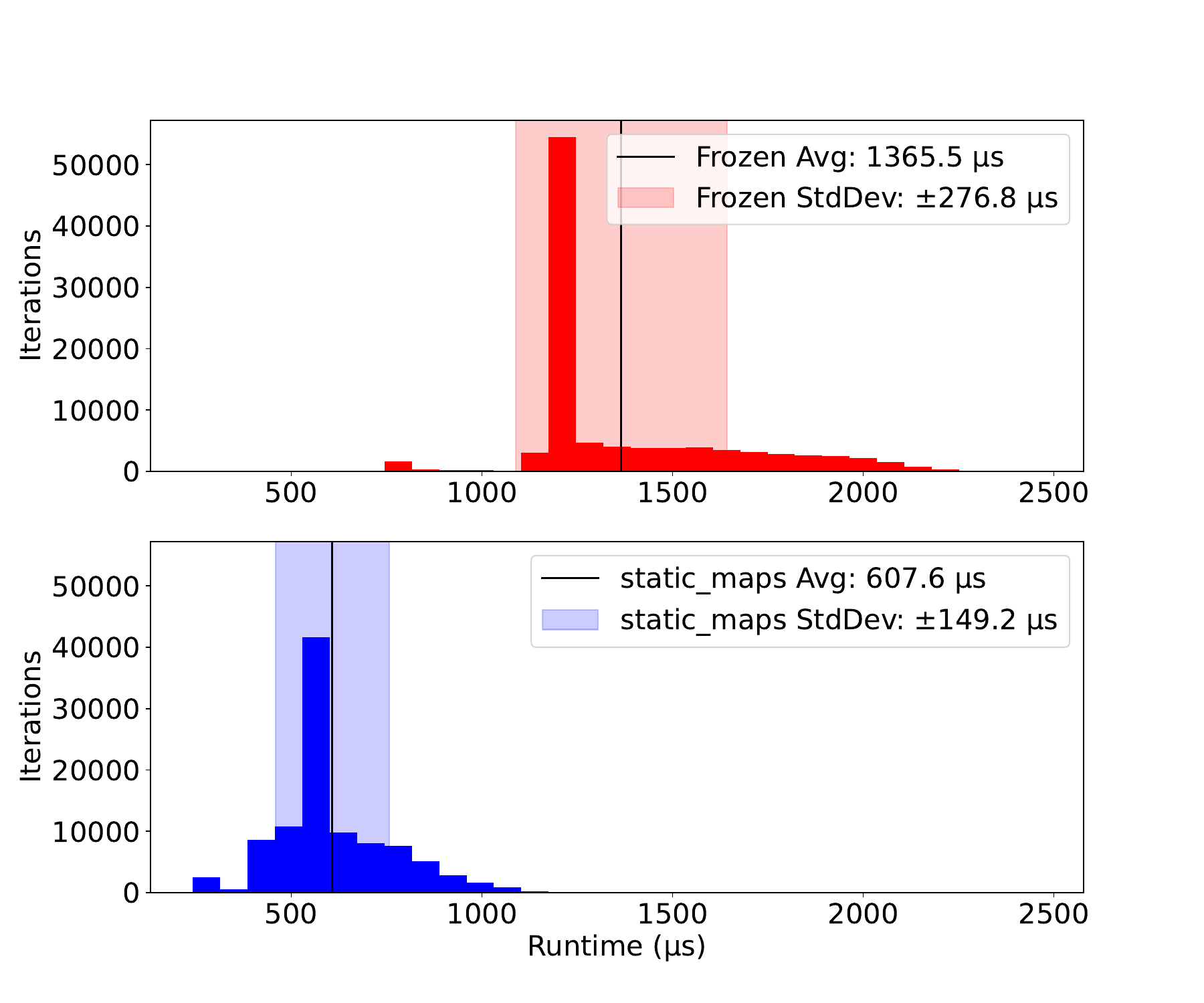} \\
\rotatebox{90}{\textbf{Stocks}} & \includegraphics[width=2.95in]{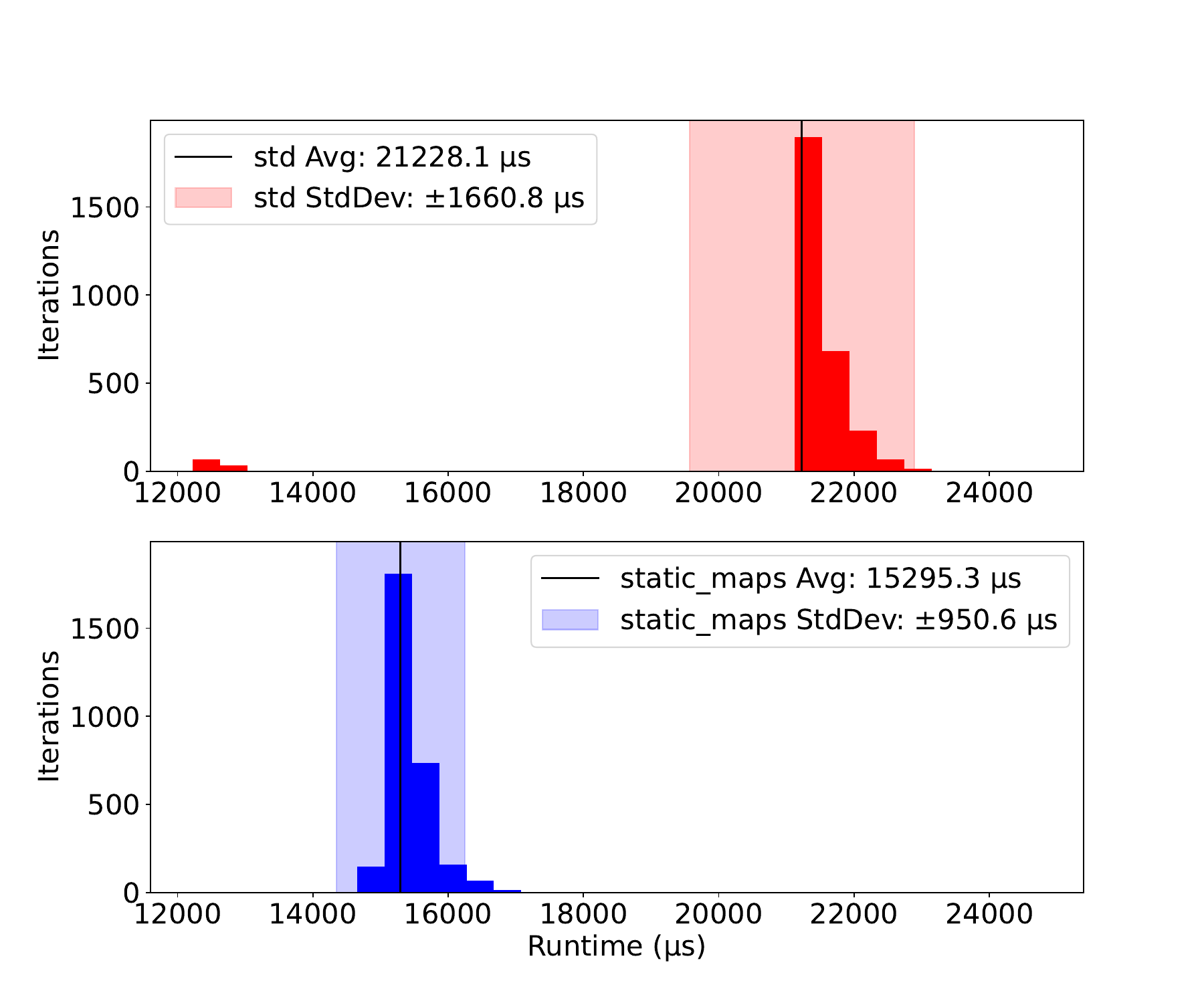} & \includegraphics[width=2.95in]{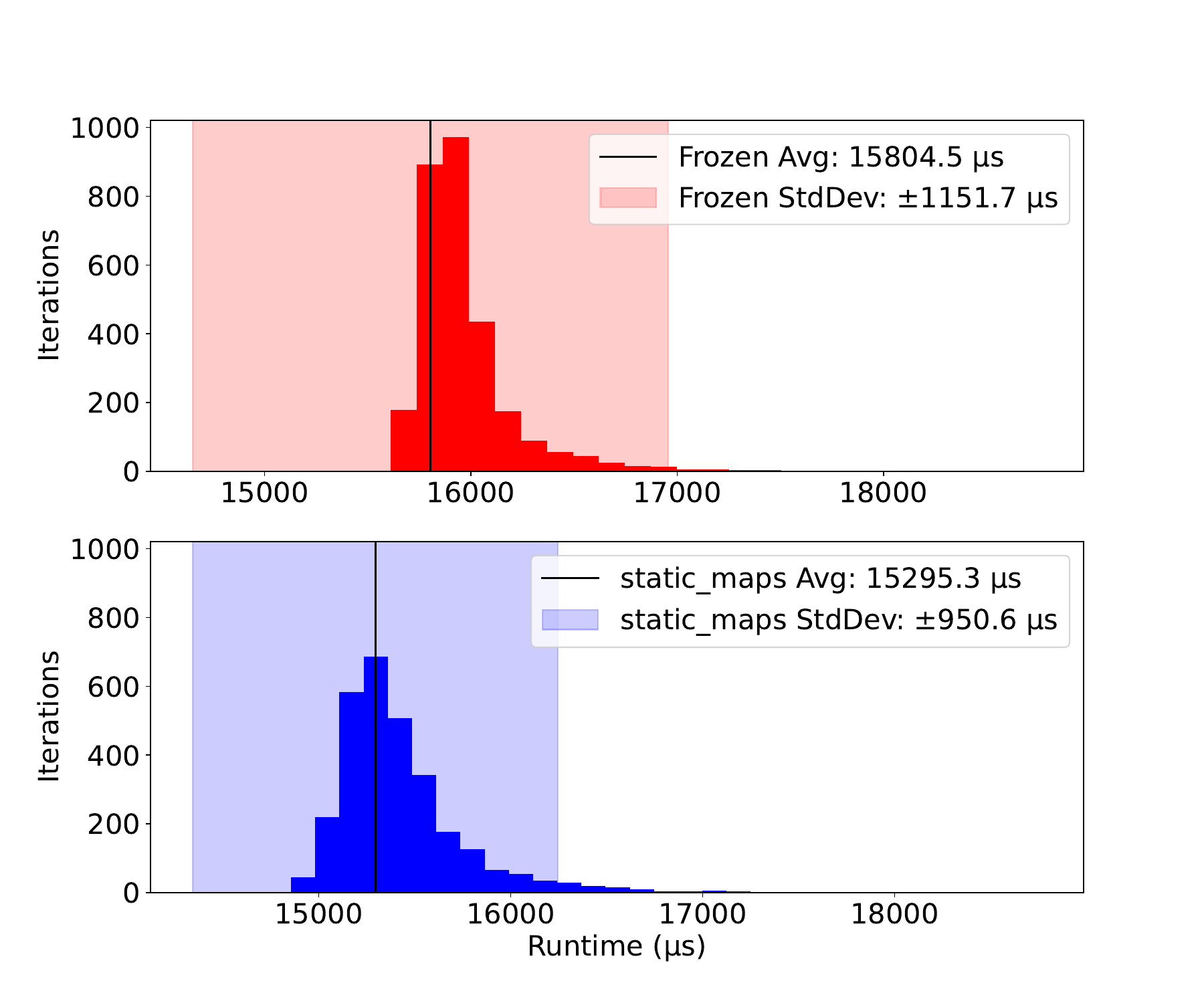}
\end{tabular}
\caption{Runtimes of construction and lookup for {\tt static\_maps::map} against {\tt std::map} on various demos. Lower is better.}
\label{fig:umap-demos}
\end{figure}

\subsection{{\tt static\_maps::unordered\_map} vs. {\tt PTHash}, {\tt gperf}}
PTHash and gperf do not provide similar interfaces to {\tt unordered\_map}, so
for convenience we only tested them on one arbitrarily picked demo. We have no
reason to believe that the other demos would produce poorer results.

\begin{figure}[H]
  \centering
\begin{tabular}{c>{\centering\arraybackslash}m{3in}>{\centering\arraybackslash}m{3in}}
  & vs. {\tt PTHash} & vs. {\tt gperf} \\
\rotatebox{90}{\textbf{Elements}} & \includegraphics[width=2.95in]{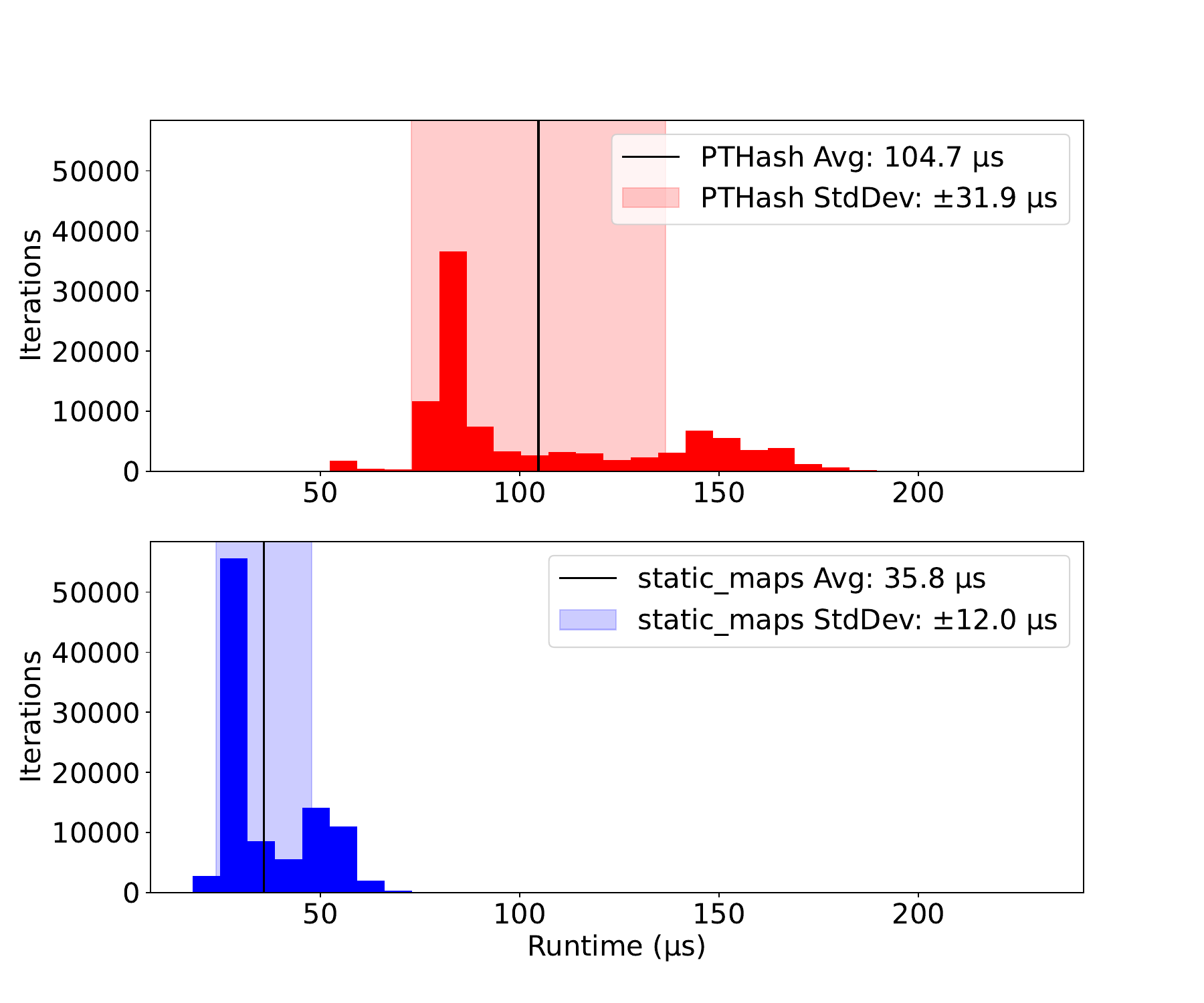} & \includegraphics[width=2.95in]{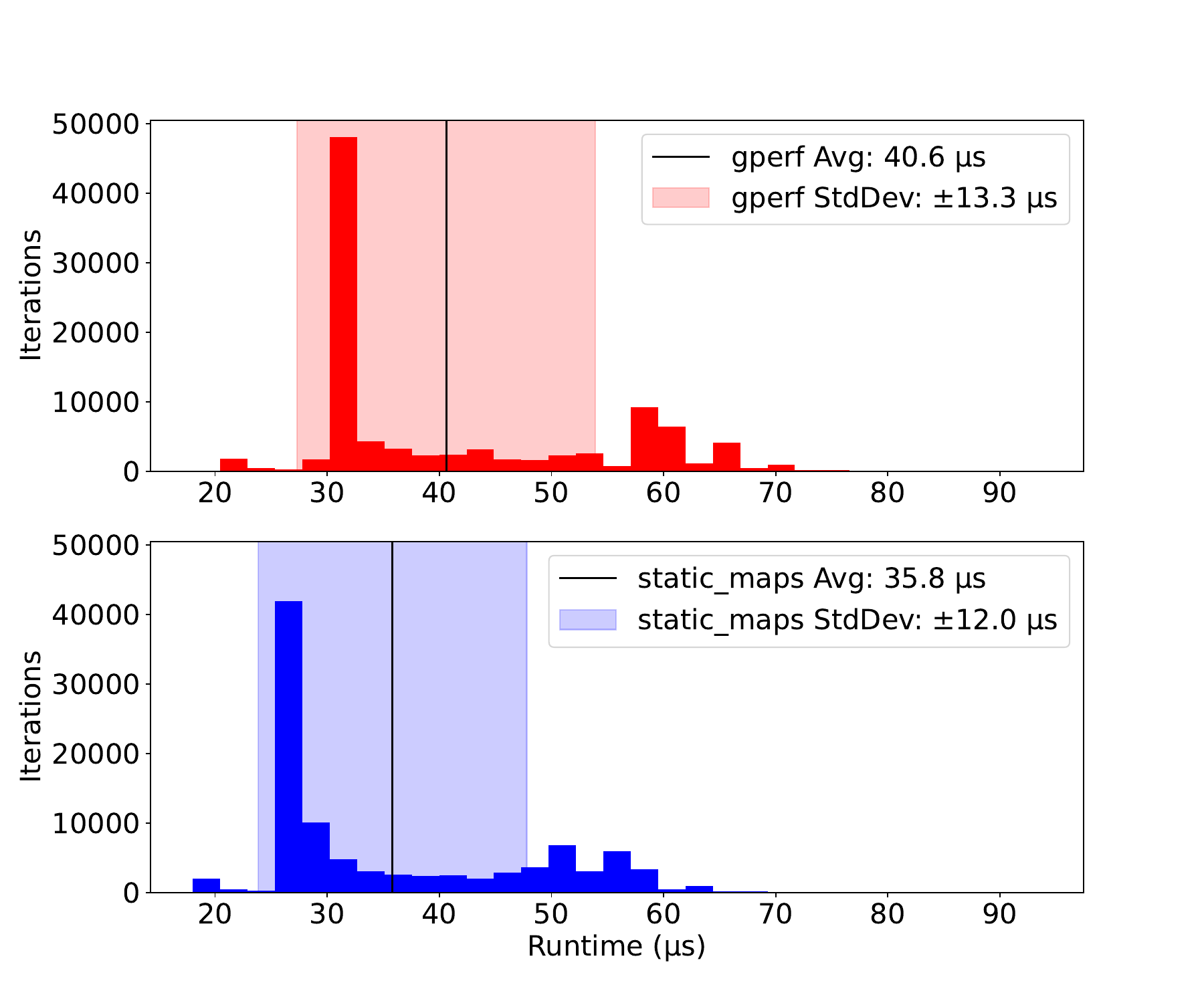}
\end{tabular}
\caption{Runtimes compared to alternatives to drop-in replacements of {\tt unordered\_map}. Lower is better.}
\label{fig:alternatives-demo}
\end{figure}

\subsection{Compilation times}

Compilation times were run on the same system as the demo runtimes. All are
averages of ten compilations in seconds.

\begin{table}[H]
  \centering
  \begin{tabular}{cccc}
    & Element Demo & Codon Demo & Stock Demo \\
    \toprule
    {\tt std::map} & 1.0 s & 0.9 s & 1.5 s \\
    \midrule
    {\tt std::unordered\_map} & 0.9 s & 0.9 s & 1.5 s \\
    \midrule
    {\tt frozen::unordered\_map} & 1.0 s & 0.9 s & 21.1 s \\
    \midrule
    {\tt static\_maps::map} & 1.3 s & 1.0 s & 10.7 s \\
    \midrule
    {\tt static\_maps::unordered\_map} & 7.3 s & 2.2 s & 55.2 s
  \end{tabular}
  \caption{Compilation times using a variety of map and hash map
    classes on various demos. Lower is better.}
\label{tab:compilation-times}
\end{table}

\section{Discussion}
A {\tt constexpr map} or {\tt unordered\_map} removes any runtime
impact of a lookup where the key is known at compile time. Both {\tt Frozen}
and {\tt static\_maps} provide this advantage. However, {\tt Frozen} uses
its own string type, necessitating explicit type changes in existing
projects to adopt a {\tt constexpr unordered\_map}. {\tt static\_maps}
supports {\tt std::string\_view} which can wrap {\tt std::string} and
{\tt char*} strings. This benefit applies to new applications of {\tt
  consteval} maps and hash maps.

\subsection{{\tt map}}
While {\tt static\_maps::map} is an alternative to {\tt std::map},
note that the compiler does optimize {\tt std::map} more effectively
in the codon demo (Figure~\ref{fig:map-demos}).

\subsection{{\tt unordered\_map}}
The demos show that {\tt constexpr} hash maps benefit runtime lookup
in a more pronounced manner. Our {\tt static\_maps::unordered\_map}
outperforms all competition to varying degrees
(Figure~\ref{fig:umap-demos} and
Figure~\ref{fig:alternatives-demo}). But overall, it appears to be
a better performer in general. The benefits are in both construction
time and time per lookup but do not diminish when lookups begin to
dominate the runtime, as in the codon to amino acid demo.

Preliminary evidence suggests read-only demos benefit more. The stocks
demo with mutable values is still significantly faster than {\tt
  std::unordered\_map}, but more similar to {\tt Frozen}. Even though
the full map is not {\tt constexpr}, shifting some of the work to
compile time and using a perfect hash still shows improvements.

By comparison to {\tt Frozen}, we also see that {\tt static\_maps} has
runtime benefits beyond those inherent to {\tt constexpr}. Our
implementation of the hash map includes optimizations like small
integer sizes, custom modulo functions, and storing values directly in
the table. This improves lookup times over the {\tt constexpr} and
{\tt constinit} maps provided by {\tt Frozen}. As lookups dominate the
runtime of the application, the benefits become more apparent (such as
in the codons demo). The time to write values remains competitive as
well, slightly faster than {\tt Frozen} on the stocks demo, but also
notably with lower variance (which is helpful for low-latency
applications such as trading). All of these demos are run with
generally applicable parameters. While we did not customize {\tt
  static\_maps} for each demo for fairness (\emph{i.e.}, using a fixed
$\delta$ for all demos), the ability to customize could improve
performance further on specific keysets. {\tt Frozen} does not provide
this kind of interface for modifying its operation.

Other perfect hashing solutions such as {\tt gperf} require more
friction to use. {\tt static\_maps} compares favorably in both
convenience and runtime to these alternatives. In our element to mass
demo, our runtime was $34\%$ of the same demo using a hash function
from {\tt PTHash}. {\tt PTHash} focuses its benchmarks on keysets of a
million keys or more, so it has separate use cases which we have not
covered. The lookups must be implemented by the user of the library,
since {\tt PTHash} just finds a hash function. Altogether, {\tt
  static\_maps} would be preferable in cases where a {\tt
  std::unordered\_map} is a component of a program and tables are not
being used as a database. On the other hand, {\tt gperf} generates
code which can be customized to use as a hash map. This command line
tool competes admirably to replace a {\tt std::unordered\_map} in
static-keyset scenarios. The resulting interface differs from {\tt
  std::unordered\_map} enough that it is difficult to switch a large
project to using {\tt gperf}, especially if the keyset may need to be
changed for future builds. {\tt static\_maps::unordered\_map} improves
over {\tt gperf} in these aspects and still has a runtime $88\%$ of an
{\tt gperf}-generated map in our element to mass demo.

Altogether, the {\tt static\_maps} library provides compelling
alternatives to the standard library's map classes for
high-performance code. In our survey of options for mapping static
keysets, it provided the fastest runtimes in each relevant scenario
and can be adapted fairly well to other scenarios. For instance, if
the number of lookups were significantly decreased, a lower $\delta$
may be preferred, decreasing construction time (which would be
weighted with relatively higher importance) at the cost of increased
secondary hashing frequency (which would be weighted with relatively
lower importance).

Compilation times are higher using our library
(Table~\ref{tab:compilation-times}). Even so, they remain fairly
modest overall. Greater use of {\tt static constexpr} qualifiers
throughout may be useful to helping the compiler memoize repeated
tasks. In non-header-only libraries, it may be possible to compile
each use a single time so the time of repeated compilation would not
be noticeable in a large project with many {\tt unordered\_map} types
instantiated. The downsides of avoiding recompilation include less
inlining unless the compiler can use Link Time Optimization.

\subsection{Future ideas}
\subsubsection{Smaller raw hashes}
The hash function that is run at every lookup includes generating a
raw hash, which outputs a 64-bit integer. For fairness, we did not
modify this behavior, even when we're truncating these values to use a
smaller {\tt SIZE\_TYPE}. A raw hash that can adapt to smaller integer
types may improve runtime further if the compiler does not already
optimize this operation.

\subsubsection{Dynamic keysets}
For dynamic use cases where new keys may be added at runtime, a
wrapper class could be devised in a vein similar to how we made the
values mutable. A hash map class could encapsulate a {\tt constexpr}
hash map queried first on lookup then revert to a dynamic hash table
for keys not known at compile time. For online applications where
regular rebuilds would update the keyset, this could achieve similar
runtime speeds to the fixed keyset situations.

\section{Availability}
\subsection{Compiler compatibility}
The library compiles under {\tt gcc} 15.2.1 with argument {\tt -std=c++23}.
Often, {\tt -fconstexpr-ops-limit} will need to be increased from its default
value ($2^{25}$ at time of writing), and for large keysets {\tt
  -fconstexpr-loop-limit} may also need to be increased (default
$2^{18}$) \cite{gcc_constexpr_limit}.

The library also compiles under clang 21.1.6, again with {\tt -std=c++23}. The
relevant limit for clang is {\tt -fconstexpr-steps}, which often needs increased
to a value above $2^{20}$.

To compile the demos in clang, {\tt -fconstexpr-steps=100000000} is a sufficient
value for all demos. The stock to price demo under clang also requires
{\tt -fexperimental-new-constant-interpreter} \cite{clang_constexpr_steps}.

Earlier versions of both these compilers have experimental support for
{\tt C++23} features, and we used some earlier versions of {\tt gcc} and
clang during development.  The precise past versions on which the
entire library compiles are not known.


\subsection{Repository of {\tt static\_maps} code}
All code from this work is available with an MIT license at
\href{https://doi.org/10.6084/m9.figshare.31416038}{10.6084/m9.figshare.31416038}

\bibliographystyle{unsrtnat}

\begin{thebibliography}{19}
\providecommand{\natexlab}[1]{#1}
\providecommand{\url}[1]{\texttt{#1}}
\expandafter\ifx\csname urlstyle\endcsname\relax
  \providecommand{\doi}[1]{doi: #1}\else
  \providecommand{\doi}{doi: \begingroup \urlstyle{rm}\Url}\fi

\bibitem[Adelson-Velsky and Landis(1962)]{adelson1962algorithm}
Georgy Adelson-Velsky and Evgenii Landis.
\newblock An algorithm for the organization of information.
\newblock \emph{Doklady Akademii Nauk SSSR}, 146\penalty0 (2):\penalty0
  263--266, 1962.

\bibitem[Bayer(1972)]{bayer1972symmetric}
Rudolf Bayer.
\newblock Symmetric binary b-trees: Data structure and maintenance algorithms.
\newblock \emph{Acta Informatica}, 1\penalty0 (4):\penalty0 290--306, 1972.

\bibitem[Guibas and Sedgewick(1978)]{guibas1978dichromatic}
Leonidas~J. Guibas and Robert Sedgewick.
\newblock A dichromatic framework for balanced trees.
\newblock In \emph{19th Annual Symposium on Foundations of Computer Science
  (SFCS 1978)}, pages 8--21. IEEE, 1978.

\bibitem[Cormen et~al.(2022)Cormen, Leiserson, Rivest, and
  Stein]{cormen2022introduction}
Thomas~H. Cormen, Charles~E. Leiserson, Ronald~L. Rivest, and Clifford Stein.
\newblock \emph{Introduction to Algorithms}.
\newblock MIT Press, 4th edition, 2022.

\bibitem[cppreference.com(2025{\natexlab{a}})]{cppref:map}
cppreference.com.
\newblock {\tt std::map} - cppreference.com.
\newblock \url{https://en.cppreference.com/w/cpp/container/map},
  2025{\natexlab{a}}.
\newblock Accessed: 2026-01-26.

\bibitem[cppreference.com(2025{\natexlab{b}})]{cppref:size_t}
cppreference.com.
\newblock {\tt std::size\_t} - cppreference.com.
\newblock \url{https://en.cppreference.com/w/cpp/types/size_t.html},
  2025{\natexlab{b}}.
\newblock Accessed: 2026-01-26.

\bibitem[cppreference.com(2025{\natexlab{c}})]{cppref:unordered_map}
cppreference.com.
\newblock {\tt std::unordered\_map} - cppreference.com.
\newblock \url{https://en.cppreference.com/w/cpp/container/unordered_map},
  2025{\natexlab{c}}.
\newblock Accessed: 2026-01-26.

\bibitem[Fredman et~al.(1984)Fredman, Komlós, and Szemerédi]{Fredmanetal1984}
Michael~L. Fredman, János Komlós, and Endre Szemerédi.
\newblock Storing a sparse table with {O}(1) worst case access time.
\newblock \emph{Journal of the ACM}, 31\penalty0 (3):\penalty0 538--544, 1984.
\newblock \doi{10.1145/828.1884}.

\bibitem[Serang(2026)]{serang:algorithms}
Oliver Serang.
\newblock Algorithms in {Python}, 2026.

\bibitem[Belazzougui et~al.(2009)Belazzougui, Botelho, and
  Dietzfelbinger]{belazzougui2009hash}
Djamal Belazzougui, Fabiano~C. Botelho, and Martin Dietzfelbinger.
\newblock Hash, displace, and compress.
\newblock In \emph{Algorithms-ESA 2009: 17th Annual European Symposium,
  Copenhagen, Denmark, September 7-9, 2009. Proceedings}, pages 682--693.
  Springer, 2009.

\bibitem[cppreference.com(2025{\natexlab{d}})]{cppref:consteval}
cppreference.com.
\newblock {\tt consteval} specifier (since {C++20}) - cppreference.com.
\newblock \url{https://en.cppreference.com/w/cpp/language/consteval.html},
  2025{\natexlab{d}}.
\newblock Accessed: 2026-01-26.

\bibitem[Grudinin(2021)]{frozen2021grudinin}
Sergei Grudinin.
\newblock Frozen: {A} header-only {C++14} library for constexpr{,} immutable
  data structures.
\newblock \url{https://github.com/serge-sans-paille/frozen}, 2021.

\bibitem[Pibiri and Trani(2021)]{pibiri2021pthash}
Giulio~Ermanno Pibiri and Roberto Trani.
\newblock {PTHash}: Revisiting {FCH} minimal perfect hashing.
\newblock In \emph{Proceedings of the 44th International ACM SIGIR Conference
  on Research and Development in Information Retrieval}, pages 1339--1348,
  2021.
\newblock \doi{10.1145/3404835.3462849}.

\bibitem[Schmidt(1990)]{schmidt1990gperf}
Douglas~C. Schmidt.
\newblock {GPERF}: A perfect hash function generator.
\newblock In \emph{Proceedings of the Second USENIX C++ Conference}, pages
  87--102, San Francisco, California, 1990.

\bibitem[Vigna(2016)]{vigna2016xorshift}
Sebastiano Vigna.
\newblock An experimental exploration of {Marsaglia}'s xorshift generators,
  scrambled.
\newblock \emph{ACM Transactions on Mathematical Software (TOMS)}, 42\penalty0
  (4):\penalty0 1--19, 2016.
\newblock \doi{10.1145/2845077}.

\bibitem[{HandWiki Contributors}(2024)]{handwiki_xorshift}
{HandWiki Contributors}.
\newblock Xorshift: {A} class of pseudorandom number generators.
\newblock \url{https://handwiki.org/wiki/Xorshift}, 2024.
\newblock Accessed: 2024-05-22.

\bibitem[Kreitzberg et~al.(2020)Kreitzberg, Pennington, Lucke, and
  Serang]{kreitzberg2020fast}
Patrick Kreitzberg, Jake Pennington, Kyle Lucke, and Oliver Serang.
\newblock Fast exact computation of the k most abundant isotope peaks with
  layer-ordered heaps.
\newblock \emph{Analytical Chemistry}, 92\penalty0 (15):\penalty0 10613--10619,
  2020.

\bibitem[{Free Software Foundation}(2024)]{gcc_constexpr_limit}
{Free Software Foundation}.
\newblock \emph{{GNU Compiler Collection (GCC) Online Documentation}: {C++}
  Dialect Options}.
\newblock Free Software Foundation, 2024.
\newblock URL
  \url{https://gcc.gnu.org/onlinedocs/gcc/C_002b_002b-Dialect-Options.html#index-fconstexpr-loop-limit}.
\newblock Accessed: 2026-02-06.

\bibitem[{The LLVM Project}(2024)]{clang_constexpr_steps}
{The LLVM Project}.
\newblock \emph{{Clang Command Line Reference}}.
\newblock The LLVM Foundation, 2024.
\newblock URL
  \url{https://clang.gnu.org/docs/ClangCommandLineReference.html#cmdoption-clang-fconstexpr-steps}.
\newblock Accessed: 2026-02-06.

\end{thebibliography}

\end{document}